\input{aipcheck}

\documentclass[
    ,final            
  ]
  {aipproc}

\layoutstyle{6x9}

\begin{document}

\title{Cosmic-Ray Signatures of Dark Matter Decay}


\author{David Tran}{
  address={Physik-Department T30d, Technische Universit\"at M\"unchen,\\
  85748 Garching, Germany}
}

\begin{abstract}
In light of recent observations of an anomalous excess of high-energy
positrons and electrons by the PAMELA and Fermi LAT experiments, 
we investigate exotic cosmic-ray signatures in scenarios with 
unstable dark matter that decays with an extremely long lifetime. 
We identify decay modes capable of explaining the observed anomalies 
and mention constraints arising from measurements of antiprotons and 
gamma rays. We also discuss complementary tests by measurements of 
anisotropies in diffuse gamma rays which should be accessible to Fermi.
\end{abstract}

\maketitle


\section{Introduction}

Although the nature of the ubiquitous dark matter particles, which make 
up a significant portion of the energy density of the Universe, remains a 
mystery, some dark matter properties have been established by indirect means. 
Among these, stability of the dark matter particles is often listed. 
However, perfect dark matter stability is merely an assumption, where we 
really only have a lower bound on the lifetime of dark matter particles 
from the fact that we observe their presence in the Universe today.
One reason for the popularity of this assumption may be that in the most 
commonly studied class of dark matter candidates, the weakly interacting 
massive particles (WIMPs, in particular the lightest neutralino), 
it is difficult to plausibly achieve a sufficiently long lifetime unless 
the decay of dark matter particles into Standard Model particles is 
completely forbidden by a symmetry.
Nevertheless, there are some well-motivated dark matter candidates which 
decay with lifetimes well in excess of the age of the Universe.

\section{Unstable Dark Matter and Indirect Detection}
Aside from WIMPs, super-weakly interacting massive particles 
(super-WIMPs such as the gravitino) are natural dark matter candidates 
with typically very long lifetimes. A moderate suppression of
their couplings is often enough to obtain a lifetime compatible with dark 
matter, without the need to impose any exact symmetries. 
However, when such unstable dark matter particles eventually decay, 
they produce ordinary Standard Model particles in the process which may be 
detectable in cosmic-ray measurements if the decay rate is sufficiently 
high. 
Suitable channels for indirect dark matter seaches include primarily 
antimatter as well as diffuse gamma rays. 

The satellite experiments PAMELA~\cite{Adriani:2008zr} and Fermi 
LAT~\cite{Abdo:2009zk} have recently measured, respectively, 
a dramatic rise in the positron fraction between 10 and 100 GeV, 
and a smooth, featureless electron-plus-positron flux at energies 
up to 1 TeV that is harder than expected. 
It is interesting to entertain the possibility that these high-energy 
positrons, left unexplained by conventional production and propagation models, 
stem from the decay or self-annihilation of dark matter particles in the 
Galactic halo. 
In contrast to the more widely studied case of dark matter annihilation, 
where the rate of cosmic rays produced is proportional to the square of the 
dark matter density, in the case of dark matter decay the source term 
is linear in the dark matter density profile. 
This leads to some important qualitative differences.
Generally, there is no enhancement of indirect signals (``boost factors'') 
resulting from substructures in regions with high dark matter concentrations. 
Furthermore, the results for cosmic-ray fluxes are more 
robust as they are less sensitive to uncertainties in the distribution 
of dark matter in the Galactic halo. 

\section{Constraints on Dark Matter Properties}

Since positrons propagating in the interstellar medium efficiently lose energy 
by synchrotron radiation and inverse Compton scattering, the high-energy 
positrons and electrons observed by PAMELA and Fermi must originate from a 
local source. 
The decay or self-annihilation of dark matter particles is a possible 
candidate, with $e^+e^-$ pair-production in the magnetic fields of nearby 
pulsars being the most popular astrophysical alternative (see \textsl{e.g.}
\cite{Hooper:2008kg}). 

Under the assumption that dark matter decay is indeed the origin of the 
high-energy positrons, we analyzed various decay modes for scalar 
and fermionic dark matter candidates in a model-independent manner in order 
to understand which dark matter properties can be inferred from these 
observations~\cite{Ibarra:2008jk}. 
We sampled various dark matter masses, while treating the dark 
matter lifetime as a free parameter. For a scalar dark matter particle, 
we examined the two-body decays 
$\phi_{\rm{DM}} \rightarrow \ell^+ \ell^-$, $Z^0 Z^0$, $W^\pm W^\mp$, 
with $\ell = $ $e$, $\mu$, $\tau$ being the charged leptons. 
The resulting injection spectra of positrons from 
fragmentation of the weak gauge bosons were obtained using the PYTHIA 6.4 
Monte Carlo code. 
For a fermionic dark matter particle, we examined the two- and three-body 
decay modes
$\psi_{\rm{DM}} \rightarrow \ell^+ \ell^- \nu$, $Z^0 \nu$, $W^\pm \ell^\mp$, 
where the three-body decays are assumed to 
be mediated by a heavy scalar particle, motivated by the interesting 
example of a hidden-sector gaugino as a dark matter candidate~
\cite{Ibarra:2009bm}. After production, charged cosmic rays undergo a 
complicated propagation 
process in the magnetic halo of our Galaxy, including diffusion on magnetic 
inhomogeneities, energy loss, convection, and annihilation in the Galactic 
disk. 
We modeled these processes by means of a semi-analytical model to propagate 
the injection spectra of positrons and antiprotons to our position in the 
Galaxy. 

\begin{figure}
  \includegraphics[height=.2\textheight]{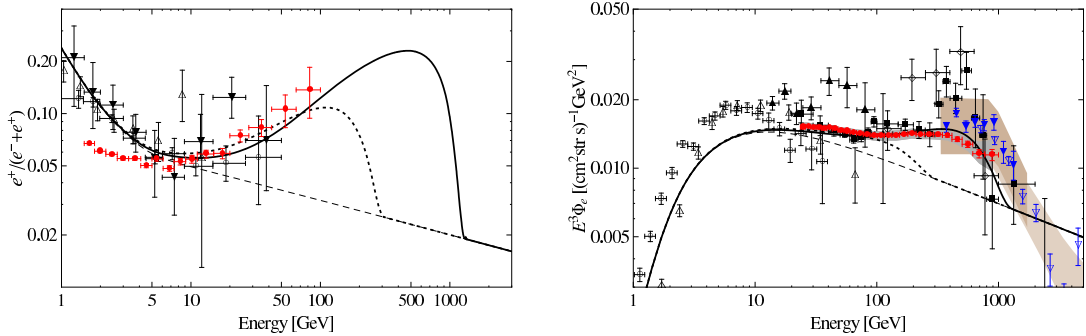}
  \caption{Scalar dark matter decaying to two muons with a lifetime of 
$\sim 10^{26}$ sec. Left panel: positron fraction, right panel: total electron-plus-positron flux. 
Dotted lines: $m_{\rm{DM}} = 600$ GeV, solid lines: $m_{\rm{DM}} = 2500$ GeV. 
The data points plotted in red are from PAMELA and Fermi LAT, respectively.}
\label{fig}
\end{figure}

The requirement that dark matter decay be able to account for the PAMELA 
anomaly yields a typical dark matter lifetime of around $10^{26}$ sec, which 
exceeds the age of the Universe by a factor $10^9$. 
Furthermore, since the rise in the positron fraction extends to at least 
100 GeV, the dark matter mass must be at least twice as large. 
The positrons generated in the fragmentation of weak gauge bosons are 
generally found to have an energy spectrum which after propagation is too 
flat to match the steep rise in the positron fraction.
Likewise, the positrons stemming from third-generation leptons in two- or 
three-body dark matter decays 
tend to produce too flat a spectrum unless the dark matter particle is 
very heavy (which seems to be in conflict with H.E.S.S. observations that seem 
to indicate a steepening of the total electron flux above 
1 TeV~\cite{Collaboration:2008aaa}). 
However, the hard leptons from two- and three-body decays of 
dark matter particles into the first two generations (or flavor-democratic 
decays into all three generations) are found to reproduce 
the rise in the positron fraction quite accurately.
Additionally taking the Fermi data into account, and requiring dark matter 
decay to explain both 
data sets simultaneously, one needs the dark matter mass to be as heavy as 
a few TeV. Furthermore, the non-observation of prominent spectral features 
in the total electron flux disfavors decays purely into the first generation 
of leptons~\cite{Ibarra:2009dr} since these typically generate distinct 
spectral features in the total electron flux. Thus, decays into the second 
generation or flavor-democratic decays are favored by the Fermi results 
(see Fig.~\ref{fig}).

One might object to dark matter masses of several TeV as being unnatural. 
A possible solution is the observation that the flux of cosmic rays from 
dark matter decay is invariant under a simultaneous rescaling of dark matter 
abundance and lifetime. This opens the possibility of having a less abundant 
population of dark matter particles which decays into the primary population 
of dark matter particles, which may be composed of stable WIMPs, such as in the 
standard neutralino dark matter scenario. Another interesting observation is 
that a dark matter lifetime of $10^{26}$ sec may be
understood in terms of Grand Unification, as a particle decaying via a 
dimension-6 operator suppressed by a mass scale $M$ has the right lifetime 
if $M$ is to be close to the GUT scale.

In addition to electron measurements, constraints arising from measurements 
of the antiproton flux and the measurement of diffuse gamma rays have to 
be taken into account. 
In particular, measurements of the antiproton-to-proton ratio by 
PAMELA~\cite{Adriani:2008zq} agree with astrophysical expectations and thus 
forbid any sizable contributions to the total antiproton flux from dark 
matter decay, strongly disfavoring hadronic decay modes. 
Furthermore, measurements of diffuse gamma rays by EGRET and Fermi LAT 
provide stringent upper bounds on gamma-ray production.
Namely, dark matter decay itself, as well as inverse Compton scattering of 
high-energy electrons and positrons produced in these decays, 
generates a contribution to the diffuse gamma-ray flux which can 
be of the same order as the extragalactic gamma-ray background for the decay 
modes and lifetimes favored by the PAMELA and Fermi observations. 
This contribution might actually be observed as a deviation from the 
expected power-law behavior of the extragalactic background. 
Another handle on indirect detection could be the observation of 
large-scale anisotropies in diffuse gamma 
rays that are expected due to the fact that we are located far from the 
center of the Galactic dark matter halo. The resulting dipole-like 
anisotropy should be accessible to Fermi LAT provided that dark matter decay is 
indeed the correct explanation for the observed cosmic-ray anomalies 
\cite{Ibarra:2009nw}.

\section{Decaying Gravitino Dark Matter}

Gravitino dark matter in models with $R$-parity violation is a 
particularly interesting example of a decaying dark matter candidate.
It has been observed that gravitino dark matter together with a high 
reheating temperature and a very small amount of $R$-parity violation can 
yield a consistent cosmology including thermal leptogenesis and primordial 
nucleosynthesis~\cite{Buchmuller:2007ui}. We restrict ourselves here to 
gravitino masses below 600 GeV, as motivated 
by universal boundary conditions at the GUT scale. 

For gravitino masses of order 100 GeV, the branching ratio for 
the $R$-parity-violating decay mode $\psi_{3/2} \rightarrow W^\pm \ell^\mp$ 
is around 50\%, with the 
hard leptons from these two-body decays providing a possible source of 
high-energy positrons.
From the model-independent considerations above we can conclude that 
the $R$-parity violation must either be flavor-blind or otherwise yield a 
rather large fraction of decays into the first two lepton generations. 
It is indeed possible to reproduce the rise in the positron fraction under 
this assumption while remaining compatible with antiproton bounds for 
certain sets of propagation parameters. 
Nevertheless, since we are only regarding gravitino masses below 600 GeV, 
one still requires astrophysical sources such as pulsars to account for 
the hard high-energy electron spectrum observed by Fermi, as these 
electrons cannot originate from the decay of such relatively 
``light'' dark matter particles~\cite{Buchmuller:2009xv}. 


\begin{theacknowledgments}
  I would like to thank W. Buchm\"uller, L. Covi, M. Grefe, A. Ringwald, 
T. Shindou, F. Takayama and in particular A. Ibarra and C. Weniger for very 
pleasant collaborations. This work was partially supported by the DFG 
cluster of excellence "Origin and Structure of the Universe.''
\end{theacknowledgments}



\bibliographystyle{aipproc}   

\bibliography{ddmbib}

\IfFileExists{\jobname.bbl}{}
 {\typeout{}
  \typeout{******************************************}
  \typeout{** Please run "bibtex \jobname" to optain}
  \typeout{** the bibliography and then re-run LaTeX}
  \typeout{** twice to fix the references!}
  \typeout{******************************************}
  \typeout{}
 }

\end{document}